\documentstyle[aps,prl]{revtex}
\draft
\begin{document}

\twocolumn[
\hsize\textwidth\columnwidth\hsize\csname @twocolumnfalse\endcsname

\title{Lifshitz-Slyozov Scaling For Late-Stage Coarsening With An
Order-Parameter-Dependent Mobility}

\author{A. J. Bray and C. L. Emmott}

\address{Theoretical Physics Group \\
Department of Physics and Astronomy \\
The University of Manchester, M13 9PL, UK}

\maketitle

\widetext
\begin{abstract}
The coarsening dynamics of the Cahn-Hilliard equation with order-parameter
dependent mobility, $\lambda(\phi) \propto (1-\phi^2)^\alpha$, is addressed
at zero temperature in the Lifshitz-Slyozov limit where the minority phase
occupies a vanishingly small volume fraction. Despite the absence of bulk
diffusion for $\alpha>0$, the mean domain size is found to grow as
$\langle R \rangle \propto t^{1/(3+\alpha)}$, due to subdiffusive transport of
the order parameter through the majority phase. The domain-size distribution
is determined explicitly for the physically relevant case $\alpha = 1$.
\end{abstract}

\bigskip
]
\narrowtext

The phenomenon of spinodal decomposition in, e.g. binary alloys, is
usually modelled by the Cahn-Hilliard equation \cite{Review}
\begin{equation}
{\partial\phi\over \partial t} = \nabla\cdot(\lambda\nabla\mu) =
\nabla\cdot \left(\lambda\,\nabla {\delta F
\over \delta \phi}\right)\ ,
\label{CH}
\end{equation}
for the order-parameter field $\phi$. Eq.\ (\ref{CH}) takes the form of
a continuity equation, $\partial_t\phi=-\nabla\cdot {\bf j}$, with
current ${\bf j} = -\lambda\nabla\mu$, where $\lambda$ is a transport
coefficient (`mobility') and the chemical potential $\mu$ is the functional
derivative, $\mu=\delta F/\delta\phi$, of a Ginzburg-Landau free energy
functional $F[\phi]$ given by
\begin{equation}
F[\phi]=\int d^dx\,\bigl({1\over 2}(\nabla\phi)^2 + V(\phi)\bigr)\ .
\label{F}
\end{equation}
Here $V(\phi)$ is the usual double-well potential whose minima (taken
here to be at $\phi=\pm 1$) represent the equilibrium phases.

In conventional treatments of (\ref{CH}), the mobility $\lambda$ is
taken to be a constant, i.e.\ independent of the order parameter $\phi$.
Recently, however, there has been considerable interest
\cite{Kitahara,Lacasta92,Lacasta93a,Lacasta93b,Yeung92,Yeung93,Zia} in
cases where $\lambda$ depends explicitly on $\phi$, notably through the
dependence $\lambda(\phi) = \lambda_0(1-\phi^2)$.
This interest has a physical origin. It has been
noticed \cite{Kitahara} that when one models the coupling to an
external driving field $E$, such as gravity, through an additional term
$F_1[\phi] = -E \int d^dx\, z\phi({\bf x})$ in $F[\phi]$ (where here
the field ${\bf E}$ is in the $z$-direction),
this extra term does not change (\ref{CH}) unless $\lambda$ depends
on $\phi$. This is because $\delta F_1/\delta\phi=-Ez$, and $\nabla^2 z=0$.
Physically, it is clear that an external field of this form accelerates
the phase separation, so $\lambda$ must be $\phi$-dependent. Indeed,
phenomenological derivations \cite{Kitahara,Langer} of $\lambda$ yield
precisely the form $\lambda \propto (1-\phi^2)$ alluded to above.
Furthermore, the coarsening dynamics of this model has been studied
by computer simulations, both with \cite{Lacasta93b,Yeung92,Zia} and
without \cite{Lacasta92} external driving forces. It is therefore
interesting to study this problem in its own right, even without external
driving forces.

In this Communication, we study a general class of systems described by
Eq.\ (\ref{CH}) with
\begin{equation}
\lambda = (1-\phi^2)^\alpha
\label{mobility}
\end{equation}
(we absorb the constant $\lambda_0$ into the timescale). To make analytical
progress, we specialize to the case where the minority phase occupies a
vanishingly small volume fraction. For the conventional case ($\alpha=0$),
this is the limit treated by the seminal work of Lifshitz and Slyozov (LS)
\cite{LS}, and by Wagner \cite{W}, which leads to the result
$\langle R \rangle \propto t^{1/3}$ for the mean domain size, and gives
an exact expression for the domain-size distribution.
For general $\alpha \ge 0$ we find
$\langle R \rangle \propto t^{1/(3+\alpha)}$. We
also determine explicitly the domain-size distribution for the physically
relevant case $\alpha=1$. (The other physically relevant case, $\alpha=0$,
has been treated by LS.)

For small volume fractions, coarsening proceeds by nucleation and growth
rather than by spinodal decomposition. For present purposes we
limit discussion to the late stages of growth, described by the LS
evaporation-condensation mechanism, by which large
domains (of radius $R>R_c(t)$)  grow at the expense of small ones
(with $R < R_c$). In this regime the droplet size distribution has the
scaling form $n(R) = R_c^{-4}\,f(R/R_c)$.

Before we proceed, two comments are in order. We have noted that
phenomenological considerations indicate that (\ref{mobility}) with
$\alpha=1$ is an appropriate form for the mobility in the Cahn-Hilliard
equation (\ref{CH}). This being the case, one may wonder about the
physical relevance of the conventional (i.e.\ with $\alpha=0$)
Cahn-Hilliard equation. The point is that thermal noise, omitted from
(\ref{CH}), reduces the magnitude of equilibrium order parameter from
unity to $\phi_0$. Since thermal fluctuations are irrelevant
on large scales \cite{Review}, however, one can continue to work with the
noise-free equation (\ref{CH}), provided that ({\em inter alia}) one uses a
renormalized potential whose minima are now at $\pm \phi_0$. The bulk
mobility then takes the value $\lambda_{bulk} = 1-\phi_0^2>0$, and
conventional LS behavior is recovered.  The relevance of $\alpha=1$ is then
limited to `deep quenches', where the effect of thermal noise is small
enough that the predicted $t^{1/4}$ growth extends over a significant time
domain (before $t^{1/3}$ LS growth eventually sets in). In simulations,
of course, one can simply work at zero temperature, when the $t^{1/4}$
behavior (or $t^{1/(3+\alpha)}$ in general) will describe the asymptotic
growth.

The second comment concerns the role of surface diffusion. It has often
been stated that (without thermal noise) a factor $(1-\phi^2)$ in the
mobility prevents bulk diffusion, and therefore surface diffusion (i.e.\
diffusion along the interfaces), which leads to $t^{1/4}$ growth, is the
dominant coarsening mechanism in this case. It is true that the bulk
diffusion constant vanishes for $\alpha>0$ [see (\ref{phitilde}) below].
In the far off-critical systems discussed here, however, where the minority
phase does not percolate, surface diffusion alone cannot lead to
large-scale coarsening. It turns out that for $\alpha>0$ there is still
bulk {\em transport}, although this of a {\em subdiffusive}, rather than
diffusive, character.

We begin by considering a single spherical domain of `plus' phase,
with radius $R$, immersed in a sea of `minus' phase. We suppose that
the minus phase is supersaturated with the plus phase, i.e.\
$\phi=-1+\epsilon$ at infinity, and we work in the limit of small
supersaturation, $\epsilon \ll 1$. First note that the chemical
potential $\mu$ is given by
\begin{equation}
\mu = {\delta F \over \delta \phi} = V'(\phi) - \nabla^2\phi\ ,
\label{chempot}
\end{equation}
where the prime indicates a derivative.
In the bulk phases, away from the interface, $\phi$ varies slowly
in space and the $\nabla^2\phi$ term in (\ref{chempot})
can be neglected. Setting $\phi=-1+\tilde{\phi}$ in the minus phase
(with $\tilde{\phi} = \epsilon$ at infinity), Eq.\ (\ref{chempot})
gives, to lowest order in $\tilde{\phi}$,
\begin{equation}
\mu(r) = V''(-1)\tilde{\phi}(r)\ , \ \ \ \ r>R\ .
\label{mu}
\end{equation}
Inserting this result into (\ref{CH}), using (\ref{mobility}) for
$\lambda$, and retaining the leading terms for $\tilde{\phi} \ll 1$,
gives, in the minus phase (away from the interface),
\begin{equation}
\partial_t\tilde{\phi} = 2^{\alpha} V''(-1)
\nabla\cdot (\tilde\phi^{\alpha}\nabla\tilde \phi)\ .
\label{phitilde}
\end{equation}
Note that, except for $\alpha=0$, this equation is not of the usual
diffusive form. We shall find, nevertheless, that it still leads to
bulk transport, albeit of a subdiffusive character.

We now make the usual assumption (to be verified {\em a posteriori})
that the interface moves slowly enough (for large $R$) that the chemical
potential is always in equilibrium with the interface. Then the
time-derivative term can be set to zero in (\ref{phitilde}).
Using the linear relation (\ref{mu}) between $\mu$ and $\tilde{\phi}$
in the bulk minus phase, (\ref{phitilde}) can be recast as
\begin{equation}
\nabla^2 (\mu^{1+\alpha}) = 0\ ,
\label{laplace}
\end{equation}
a simple generalization of the Laplace equation $\nabla^2 \mu=0$ obtained
when $\alpha=0$.

What are the boundary conditions on (\ref{laplace})? At infinity, we have
\begin{equation}
\mu(\infty) \equiv \mu_\infty = V''(-1)\tilde{\phi}(\infty)
= V''(-1)\epsilon\ .
\label{bc1}
\end{equation}
The second boundary condition, at $r=R$, is just the usual
Gibbs-Thomson boundary condition
\begin{equation}
\mu(R) = \sigma/R\ ,
\label{GT}
\end{equation}
where $\sigma$ is the surface tension. To derive (\ref{GT}), one first writes
(exploiting the spherical symmetry) $\nabla^2\phi = \partial_r^2\phi
+(2/r)\partial_r\phi$ in (\ref{chempot}). Then one multiplies
(\ref{chempot}) by $\partial_r\phi$ and integrates across the interface.
Using the fact that $\partial_r\phi$ is sharply peaked at the interface
gives $\mu(R)\Delta\phi = -(2/R)\int dr\,(\partial_r\phi)^2 = -2\sigma/R$,
where $\Delta\phi=-2$ is the discontinuity in $\phi$ across the interface.
This reproduces (\ref{GT}).

The solution of (\ref{laplace}) with boundary conditions (\ref{bc1}) and
(\ref{GT}) is
\begin{equation}
\mu^{1+\alpha}(r) = \mu_{\infty}^{1+\alpha} + \left[\left(\frac{\sigma}{R}
\right)^{1+\alpha} - \mu_{\infty}^{1+\alpha}\right]\frac{R}{r}\ ,\ \ \
r \ge R\ .
\end{equation}
The time-dependence of $R$ is obtained by considering the flux of material
to (or from) infinity. The current $j$ through the minus phase is
\begin{eqnarray}
j(r) & = & -\lambda\partial_r\mu = -(2\tilde{\phi})^\alpha\partial_r\mu
\nonumber \\
 & = & \frac{2^\alpha}{(1+\alpha)[V''(-1)]^\alpha}\,
\left[\left(\frac{\sigma}{R}\right)^{1+\alpha} -
\mu_{\infty}^{1+\alpha}\right]\frac{R}{r^2}\ ,
\end{eqnarray}
leading to an outward flux of material $f = -2\times 4\pi R^2\,dR/dt
= 4\pi r^2 j(r)$ (where the factor of 2 on the left represents the
difference of $\phi$ between the two phases). This gives
\begin{equation}
\frac{dR}{dt} = \frac{C}{R}\left(\frac{1}{R_c^{1+\alpha}}
- \frac{1}{R^{1+\alpha}}\right)\ ,
\label{v}
\end{equation}
where $C=[2\sigma/V''(-1)]^\alpha[\sigma/2(1+\alpha)]$ is a constant,
and $R_c = \sigma/\mu_\infty = \sigma/\epsilon V''(-1)$ is the critical
radius, i.e.\ the domain will grow if $R>R_c$ and shrink if $R<R_c$.

For the case of zero supersaturation ($\epsilon=0$), $R_c=\infty$ and
$dR/dt = -C/R^{2+\alpha}$. In this case all drops shrink (by evaporation
of material to infinity). The collapse time $t_c$ for a drop of initial
size $R$ scales as $t_c \propto R^{3+\alpha}$, which already suggests
the scaling $R_c(t) \sim t^{1/(3+\alpha)}$ when evaporation and condensation
mechanisms compete in the many-domain situation.

Consider now a dilute assembly a spherical drops of various sizes. The
derivation, for general $\alpha$, of the scaling distribution of sizes
follows that of LS for $\alpha=0$. The basic idea is that one has a
time-dependent critical radius $R_c(t)$ which is determined
self-consistently. Suppose that, in the late stages of growth, the
distribution of domain radii is given by the scaling form
\begin{equation}
n(R,t) = {1\over R_c^4} f\left({R\over R_c}\right)\ ,
\label{bno}
\end{equation}
where $n(R,t)dR$ is the number of domains per unit volume with radii
in the interval $(R,R+dR)$. The prefactor $R_c^{-4}$ ensures that the
volume fraction occupied by the domains,
$\psi = \int dR\,(4\pi R^3/3)n(R,t)$, is conserved.
Inserting (\ref{bno}) into the continuity equation
\begin{equation}
{\partial n \over \partial t} + {\partial \over \partial R}(v n) =0,
\end{equation}
where $v=dR/dt$ is the domain-wall velocity given by (\ref{v}), yields
$$
{\dot R_c \over R_c^5} \left( 4f + xf' \right) =
$$
\begin{equation}
{ C \over R_c^{7+\alpha} } \left[\left ({1\over x} -
{1\over x^{2+\alpha}}\right)f'
 + \left({2+\alpha \over x^{3+\alpha}} - {1\over x^2}\right)f \right]\ ,
 \label{feq}
\end{equation}
where $\dot{R_c}\equiv dR_c/dt$.
Consistency requires that the $R_c$ dependence drop out of this equation.
This gives
\begin{equation}
R_c^{2+\alpha}\,\dot{R_c}  = \gamma C\ ,
\label{16}
\end{equation}
where $\gamma$ is a constant, implying
\begin{equation}
R_c(t) = [\,(3+\alpha)\,\gamma C t\,]^{1/(3+\alpha)}\ .
\label{Rc}
\end{equation}
Thus the characteristic domain size grows as $t^{1/(3+\alpha)}$ as
anticipated. This result generalizes the usual $t^{1/3}$ LS
growth law.

Using (\ref{Rc}) in (\ref{feq}), the latter can be integrated directly,
in the form
\begin{equation}
\int {df\over f} = \int \frac{dx}{x}\,
{2+\alpha - x^{1+\alpha } - 4\gamma x^{3+\alpha}
	\over \gamma x^{3+\alpha} - x^{1+\alpha} + 1 }
\label{fint}
\end{equation}
where we remind the reader that $x$ is the scaled radius, $x=R/R_c$.

It would seem that there is a family of solutions, parameterized by
$\gamma$. In fact this is not so -- there is  a unique value of gamma,
determined following the method used by LS \cite{LS}. First we recall
that conservation of the order parameter requires that the total
volume of the domains in the late-stage scaling regime (where the
value of the order parameter in the majority phase approaches $-1$)
be conserved, i.e.\
\begin{equation}
\frac{4\pi}{3}\int_0^\infty dR\,R^3\,n(R,t)
= \frac{4\pi}{3}\int_0^\infty dx\,x^3f(x) = \psi,
\label{norm}
\end{equation}
where $\psi$ is the volume fraction of the minority phase.
It follows that there is a maximum value, $x_m$,
of $x$ above which $f$ must vanish. Otherwise, (\ref{fint}) implies
$f \sim x^{-4}$ for $x \to \infty$, and the integral (\ref{norm}) will
be (logarithmically) divergent. In fact, the denominator of the integral
in (\ref{fint}) must have a double zero at $x=x_m$.  To see why this is so,
consider the time evolution, for a given domain, of the scaled radius
$x=R/R_c$. From (\ref{v}) one obtains
\begin{eqnarray}
\dot x &=& { \dot R \over R_c} - {R \dot R_c \over R_c^2 }\\
       &=& {1\over (3+\alpha)\gamma t } \left( {1\over x} - {1\over
				x^{2+\alpha}} - \gamma x \right)  \\
       &\equiv& {1\over (3+\alpha)\gamma t }\,g(x)\ ,
\label{flow}
\end{eqnarray}
the last equation defining $g(x)$.

The function $g(x)$ has a single maximum in the interval $(0,\infty)$.
If $\gamma < \gamma_0$, where
\begin{equation}
\gamma_0 = \left(\frac{1+\alpha}{3+\alpha}\right)\,
\left(\frac{2}{3+\alpha}\right)^{2/(1+\alpha)}\ ,
\label{g0}
\end{equation}
then the maximum lies above the $x$-axis and $g(x)$ has two zeros $x_1$ and
$x_2$, with $x_1<x_2$.  Under the dynamics (\ref{flow}), if $x<x_1$
initially,  then $x$ flows to zero, whereas if $x>x_1$ then $x$ flows to
$x_2$. However, as $t\to\infty$, $x_2R_c \to \infty$ which violates
the conservation of the order parameter. Similarly, if $\gamma>\gamma_0$,
$g(x)$ is negative everywhere in $(0,\infty)$, and all domains flow to
zero size, which again violates the conservation. We conclude that
$\gamma=\gamma_0$

For $\gamma = \gamma_0$, the maximum of $g(x)$ lies on the $x$ axis, and
$g(x)$ has a double zero at
\begin{equation}
x_m = \biggl({3+\alpha \over 2}\biggr)^{1/(1+\alpha)}\ .
\end{equation}
Eq.\ (\ref{fint}) then gives
\begin{eqnarray}
\ln f(x)& =& \int {2+\alpha - x^{\alpha + 1} - 4\gamma_0 x^{3+\alpha}
	\over x(\gamma_0 x^{3+\alpha} - x^{1+\alpha} + 1 )}\,dx\ ,
\ \ \ x<x_m\ , \\
f(x) &=& 0\ ,\ \ \ \ \ \mbox{otherwise.}
\end{eqnarray}
While this integral cannot be evaluated for general $\alpha$, the
scaling function can be determined for the two cases of greatest
physical interest.
For $\alpha =0$,
\begin{equation}
f(x) = a_0\psi\,x^2\,(3-2x)^{-11/3}\,(x+3)^{-7/3}\,
\exp\left(-\frac{3}{3-2x}\right)\ ,
\label{alpha0}
\end{equation}
for $x<x_m=3/2$, and the usual LS result \cite{LS} is recovered.
The normalization condition (\ref{norm}) gives $a_0 = 186.13\ldots$.

For $\alpha =1$,
\begin{equation}
f(x) = a_1\psi\,x^3\,(2-x^2)^{-7/2}\exp\left(-\frac{3}{2-x^2}\right)\ ,
\label{alpha1}
\end{equation}
for $x<\sqrt{2}$, where $a_1 = 7.785\ldots$. Eq.\ (\ref{alpha1}) gives
the scaling distribution of domain sizes for the modified Cahn-Hilliard
equation simulated in reference \cite{Lacasta92},
although in that work the domain-size distribution was not measured.

It remains to justify the claim that one need consider only
stationary solutions of (6), i.e.\ that the interfaces move so slowly
that they can be regarded as stationary while $\tilde{\phi}$ relaxes.
In other words, we want to justify the `adiabatic' approximation of
treating $\tilde{\phi}$ as given by its equilibrium configuration for
the instantaneous positions of the interfaces. We consider late-stage
coarsening, when $R_c$ is the only characteristic length scale.
Using  $\tilde{\phi} =\mu/V''(-1) \sim \sigma/R_cV''(-1)$, Eq.\ (6)
gives the relaxation time of $\tilde{\phi}$, for fixed interfaces,
as $t_{rel} \sim R_c^2/V''(-1)\tilde{\phi}^\alpha
\sim R_c^{2+\alpha} V''(-1)^{\alpha-1}/\sigma^\alpha$.
{}From (\ref{16}), the characteristic interface velocity is
$\partial_t R_c \sim C/R_c^{2+\alpha}$, where $C$ is given below
Eq.\ (\ref{v}). This gives the typical distance moved by an interface in
time $t_{rel}$ as $t_{rel}\,\partial_t R_c \sim \sigma/V''(-1)$. This fixed
length (of order the interface thickness) is negligible compared to $R_c$,
justifying the approximation.

We note that the derivation of the $t^{1/(3+\alpha)}$ growth requires
that the potential $V(\phi)$ have quadratic minima, since the
amplitude of the power-law growth depends (except for $\alpha=0$) on
$V''(-1)$ through the constant $C$ in (\ref{Rc}). It is interesting that
for $\alpha=0$, $R_c(t)$ depends on the potential only through the surface
tension $\sigma$, and is therefore independent of the detailed form of
$V(\phi)$.  The question of the appropriate form of the potential for deep
quenches deserves further consideration.

We stress that the results presented here are, like the original LS
calculation, valid only in the limit where the minority phase occupies an
infinitesimal volume fraction. In particular, we anticipate significant
corrections to the domain size {\em scaling function} (\ref{alpha1}) even
for $\psi$ as small as $10^{-2}$. This is certainly so for
$\alpha=0$, and improved forms for $f(x)$ have been suggested by a number
of authors \cite{Improved}. It would be interesting to see whether
similar techniques can be used for general $\alpha$.

By contrast, we expect the domain size {\em growth law},
$\langle R \rangle \sim t^{1/(3+\alpha)}$ to hold whenever the minority
phase consists of isolated domains. Lacasta et al.\ \cite{Lacasta93a}
measure an effective growth exponent of $0.20 \pm 0.01$ from
{\em two-dimensional} simulations with $\alpha=1$ and $\psi=0.3$.
The extrapolation to late times (their Figure 3) required for comparison
with the present predictions does not, however, seem completely
straightforward. Also it should be noted that the present theory
is restricted to three dimensions. While extension to general $d>2$ is
straightforward, and will not change the growth law, the case $d=2$ is
special (because of the singular form of the Green's function for the
Laplacian) and requires a separate study \cite{Rogers}.

If both phases are continuous, one needs to consider the competing
surface diffusion process, which leads to $t^{1/4}$ growth for the
characteristic length scale (defined now,
for example, by the first zero of the pair correlation function).
We conclude that, when both phases percolate, bulk transport
($t^{1/(3+\alpha)}$ growth) dominates for $\alpha <1$, surface diffusion
($t^{1/4}$ growth) dominates for $\alpha>1$, while both processes
contribute the same $t^{1/4}$ growth for $\alpha=1$.  For $\psi=0.5$
(and $\alpha=1$) Lacasta et al.\ \cite{Lacasta93a} measure an effective
growth exponent of $0.22 \pm 0.01$, again below the expected value of
$1/4$. Longer runs may be helpful in clarifying whether this discrepancy
is real.

In summary, the Lifshitz-Slyozov theory of late-stage coarsening has been
generalized to a class of models with vanishing bulk mobility. Coarsening
occurs by subdiffusive transport of the order parameter through the
majority phase. The growth law for the mean domain size, and the scaling
form for the domain-size distribution, have been determined in the LS
limit where the minority phase occupies a vanishingly small volume
fraction. The result for the growth law, however, should hold whenever the
minority phase consists of isolated domains.

AB thanks Sanjay Puri, Andrew Rutenberg and Chuck Yeung for discussions,
and the Isaac Newton Institute, Cambridge, for hospitality during the
early stages of this work.  CE thanks EPSRC (UK) for support.

\end{document}